\documentstyle[12pt,aasms]{article}

\def\fun#1#2{\lower0.837ex\vbox{\baselineskip0ex\lineskip0.209ex
  \ialign{$\mathsurround=0ex#1\hfil##\hfil$\crcr#2\crcr\sim\crcr}}}

\def\msun{M_\odot}

\def\sles{\lower2pt\hbox{$\buildrel {\scriptstyle <}
   \over {\scriptstyle\sim}$}}
 
\def\sgreat{\lower2pt\hbox{$\buildrel {\scriptstyle >}
   \over {\scriptstyle\sim}$}}

\begin{document}

 \title{ The 2001 Superoutburst of WZ Sagittae:
   A Clue to the Dynamics of Accretion Disks}

 \vskip 1truein

 \author{ John K. Cannizzo    }
 \affil{e-mail: cannizzo@stars.gsfc.nasa.gov}
 \affil{NASA/GSFC/Laboratory for High Energy Astrophysics, 
 Emergent Information Technologies, Inc.,
 Code 662, Greenbelt, MD 20771}
 \authoraddr{NASA/GSFC/Laboratory for High Energy Astrophysics, 
 Emergent Information Technologies, Inc.,
 Code 662, Greenbelt, MD 20771}

 \vskip 3truein

 \centerline{to appear in the Astrophysical Journal (Letters)} 

\received{ 2001 September 7}
\accepted{ 2001 October 1}

\begin{abstract}

We examine the light curve
of the July-August 2001 superoutburst of WZ Sagittae.
During the decline from maximum light the
     locally defined decay time
increases from $\sim 4 $   d mag$^{-1}$ 
            to $\sim 12$ d mag$^{-1}$
over the first $\sim15$ d of the $\sim25$ d superoutburst,
as the system faded from $m_V\simeq8.5$ to $m_V\simeq10$.
 The superoutburst is caused by the
sudden accretion of $\sim10^{24}$ g
of gas onto the white dwarf, and the
     deviation from  exponentiality
in the decay light curve
is expected qualitatively during a 
``viscous decay''
     in which the dominant mode of
     depletion 
  of the gas stored
    in the accretion disk
is accretion onto the central object.
   In other words, as the mass of the accretion
disk decreases, the viscous time scale increases.
We show that the
     data are also quantitatively consistent
    with the theoretical  viscous decay time,
both calculated via a simple scaling and
also from  time dependent calculations,
when one adopts standard model parameters
for WZ Sge.

\end{abstract}

\medskip
\medskip

{\it  Subject headings:}
accretion,  accretion disks $-$ binaries: close  $-$
                                   stars: individual (WZ Sge, A0620-00)

\section{ Introduction }

WZ Sagittae is a fascinating source
which has been the subject of much research
in the field of cataclysmic variables (CVs)
   $-$
interacting binaries in which a Roche lobe filling
K or M dwarf secondary transfers matter
onto a white dwarf (WD) primary (Warner 1987, 1995). 
  WZ Sge is a member of the  dwarf novae $-$
a subclass of the CVs characterized by semiperiodic
outbursts $-$
and furthermore it is the prototype of the ``WZ Sagittae''
stars which  have short orbital periods ($P_{\rm orb}=81$ min
for WZ Sge $-$ Krzemi\'nski 1962)
and only exhibit very infrequent ``superoutbursts''
that last $\sim30$ d.
Previous superoutbursts from WZ Sge occurred in 1913, 1946,
and 1978. The  most recent  superoutburst
began 23 July 2001 and ended 18 Aug 2001.
The time since the previous superoutburst (23 yr)
is significantly less than the previous two recurrence
times (both 33 yr).

Due to the widespread interest in WZ Sge
and fortuitous timing of the 2001 superoutburst,
a large number of
amateur astronomers observed the system,
and as a result the outburst light curve
is particularly well defined.
The obvious deviation from an exponential shape,
which is rendered observable due to a combination
of both the relatively large dynamic range
in $m_V$ and also the long duration of a superoutburst,
has implications for the physics of accretion of the
gas in the accretion disk onto the WD.

Smak (1993) summarizes previous observational
and theoretical work on WZ Sge.
He discusses the superoutbursts in WZ Sge
in terms of the accretion disk limit cycle
model (Meyer \& Meyer-Hofmeister 1981,
 Cannizzo 1993a),
     and estimates an  accretion disk
mass needed to power the superoutburst of $\sim10^{24}$ g.
   There has been some debate on the quiescent evolution
   of the accretion disk in the WZ Sge stars.
  If the matter accumulates in the usual way during
quiescence,
  Smak (1993)
    calculates that the quiescent value of the
Shakura \& Sunyaev (1973) alpha parameter $\alpha_{\rm cold}$
must be smaller by a factor $\sim 10^2 - 10^3$ than for  normal
dwarf novae
   to accumulate this much gas.
  Hameury, Lasota, \& Hur\'e 
      (1997,
    see also Lasota, Kuulkers, \& Charles 1999)
     present an
  alternate viewpoint in which outbursts are triggered
  by a fluctuation in the mass transfer from the secondary
  star, and the ``normal''  $\alpha_{\rm cold}$
  value
   of a few hundredths
  prevails in quiescence.
  If the matter is brought over from the secondary in $\sim1$ d,
  for example, this would imply a temporary mass transfer
  rate of $\sim10^{19}$ g s$^{-1}$.
   Meyer-Hofmeister, Meyer, \& Liu (1998) 
     present a time dependent model using a small
$\alpha_{\rm cold}\sim 10^{-3}$,
plus evaporation of matter from the inner disk
to prevent inside-out outbursts.
In connection with very evolved systems such as WZ Sge,
Meyer \& Meyer-Hofmeister (1999)
 discuss a possible physical mechanism for generating
low $\alpha_{\rm cold}$
 in cool accretion disks with little or no partial
ionization, 
 which 
    involves a much-reduced efficiency for the
magneto-rotational instability for turbulent angular
momentum transport in the accretion disk
       (Balbus \& Hawley 1991, 1998,
 Hawley \& Balbus 1991,  see also
Gammie \& Menou 1998, Menou 2000, Balbus \& Terquem 2001).
  In this work we restrict our attention
to the outbursting state.

\section { Background } 
 
In any time dependent
physical system where processes
which operate over a wide range
in time scale
mediate the equilibrium level of a given
physical quantity $-$
in this instance the mass of the accretion
disk $-$
the controlling or governing time scale
is that which is the slowest.
For our study this is the
viscous time at the outer disk edge.
Aside from complications
such as the precessing, eccentric outer
disk which is thought to develop
and   give rise to superhumps
in the light curve (Whitehurst 1988, Hirose \& Osaki 1990,
Murray \& Armitage 1998),
the superoutbursting disk
is   relatively simple 
in that the surface density $\Sigma(r)$
at all radii far exceeds the critical surface
density $\Sigma_{\rm min}(r)$ below
which the cooling thermal transition
is thought to occur.
Therefore the entire disk
is in the hot, ionized state,
and the disk mass can only decrease
through accretion onto
the central WD.  (There
is also some
mass loss from the outer edge
due to the tidal action from 
    the secondary star,
and 
   from intermediate radii through winds.)

One can obtain a simple estimate of the
viscous time at the outer disk edge
by using
the  ``vertically-averaged''
equations which give the
radial disk structure (Shakura \& Sunyaev 1973).
Since the observed superoutburst energy
constrains the  $t=0$ mass of the disk (Smak 1993),
we require scalings of the disk variables
in terms of surface density $\Sigma$ rather than the
usual accretion rate ${\dot M}$.
Eqn. [A6] of Cannizzo \& Reiff (1992)
gives the midplane disk temperature $T=T(\Omega,\alpha_{\rm hot},\Sigma)$,
where the Keplerian frequency $\Omega(r) = (GM_{\rm WD} r^{-3})^{1/2}$
and $\alpha_{\rm hot}$ is the value of alpha in the outbursting
disk ($\simeq 0.1-0.2$, Smak 1984, 1998, 1999).
If we adopt an opacity law for the  
     ionized state $\kappa = 2.8\times 10^{24}$
g cm$^{-2}$ $\rho T^{-3.5}$
and use the condition of local hydrostatic equilibrium
$\Omega^2 h^2 = c_S^2 =  R T \mu^{-1} $ (where
$c_S$ is the local sound speed,  $h$ is the local
disk semi-thickness, $R$ the ideal  gas constant, and $\mu$
the mean molecular weight  $=0.67$),
we then get for the viscous time

$\tau_v \equiv (\alpha\Omega)^{-1} (r/h)^2 = 5.7$ d 
$m_1^{5/14}$  $r_{10}^{13/14}$  $(\alpha_{\rm hot}/0.1)^{-8/7}$
$(\Sigma/2\times 10^3$ g cm$^{-2})^{-3/7}$,

\noindent where $m_1=M_{\rm WD}/\msun$
and $r_{10} = r_{\rm outer}/(10^{10}$ cm).
The value $2\times 10^3$ g cm$^{-2}$
used in the 
 $\Sigma$ scaling comes from Smak's (1993)
estimate
of the mass $\Delta M = 10^{24}$ g in the disk
at $t=0$,
taken
     to be distributed in a steady state profile
$\Sigma(r) \propto r^{-3/4}$
 extending to $10^{10}$ cm.
The other parameters have been scaled 
  in rough  accord
with Smak's summary: 
primary mass $0.45\pm0.2\msun$
and outer disk radius $(1.1\pm0.2)\times 10^{10}$ cm.
The viscous time scale $\tau_v$
represents an $e-$folding time over which
    local surface density
 perturbations at a given radius
are smoothed out in a dynamically
evolving disk.

Assuming that the mass in the decaying disk 
  is continually redistributed so that it
always extends out to the (formal) outer disk radius,
and that $\alpha_{\rm hot}$ is constant,
we see that basic disk theory  predicts
 $\tau_v = \tau_0 (\Sigma/\Sigma_0)^{-3/7}$.
As the mass drains onto the WD
and the surface density in the disk decreases,
one should see a corresponding increase
in the locally defined time scale associated with the
decay 
   light curve at a given point, given that the
luminosity varies as
the rate of change of disk mass,
  or 
      $\partial\Sigma/\partial t$.
  Even if the global decay of the light curve
does not have precisely an exponential form,
one can always define a small local patch
of it to be exponential, so that by definition
 the time scale defined from a local
tangent to the light curve
$\tau_{\rm e-fold}(\Sigma) = \tau_{\rm e-fold}
(\partial\Sigma/\partial t)$.
Standard values of $\alpha_{\rm hot}\simeq 0.1-0.2$
imply a critical surface density for cooling
$\Sigma_{\rm min}$
at $r_{10}\sim1$ of $\sim100$ g cm$^{-2}$.
 At this point the cooling front forms
    at the
outer edge and the subsequent
decay is much faster.
Therefore,
     theory predicts that the superoutburst
should end when the initial value
of $\Sigma(r_{\rm outer}) \simeq 2000$ g cm$^{-2}$
has decreased to $\sim100$ g cm$^{-2}$,
a dynamic range of about a factor of 20.
Since $\tau_v \propto \Sigma^{-3/7}$, 
  this implies an increase by a factor $\sim3-4$
 in the locally defined decay time scale
for the superoutbursting light curve
from start to end.

\section { Data and Modeling } 
\subsection { Data }

 Figure 1 shows the $m_V$ data from the VSNET
website
of the July-Aug 2001 superoutburst of WZ Sge.
   The data are contributed by amateur
astronomers spanning many organizations
across the world.
      The first panel  presents all the individual
data points.
One can see by eye a noticeable departure
from exponentiality in the decay 
          (i.e., from a straight line
when plotted as $m_V$ versus time).
The second panel shows
   moving two day averages of 
the light curve 
binned into 0.2 d bins.
The third panel shows
   the decay time scale computed from the
light curve in the second panel.
%
From $\sim25$ July 2001 to $\sim8$ Aug 2001,
the locally defined decay time scale increased
from $\sim4$ d mag$^{-1}$ to $\sim12$ d mag$^{-1}$.
This is 
    consistent with our  scaling law from the previous
section.
(One $e-$folding [i.e., $2.718\times$] exceeds
     1 mag [i.e., $2.512\times$]
 by about 8\%,
   a 
  difference 
 which we ignore
given the crudeness of the theoretical
$\tau_v$ scaling.)


\subsection { Time Dependent Modeling}

As a final check, we utilize
our time dependent accretion disk code
to calculate a detailed light curve
(for descriptions of the code see
Cannizzo 1993b, 2001).
  We assume $M_{\rm WD} = 0.45\msun$,
$r_{\rm inner} = 9\times 10^8$ cm, and
$r_{\rm outer} = 1\times 10^{10}$ cm.
At $t=0$ we place a cold torus of gas at $0.8r_{\rm outer}$
in the form of a Gaussian, 
with a width $r_{\rm FWHM} = 1\times 10^9$ cm.
We perform two trials: one with $\Sigma_{\rm peak} = 2\times 10^4$
 g cm$^{-2}$ (i.e., $\Delta M_{\rm disk} = 1\times 10^{24}$ g), 
and one with $\Sigma_{\rm peak} = 6\times 10^4$ 
 g cm$^{-2}$ (i.e., $\Delta M_{\rm disk} = 3\times 10^{24}$ g).
This is motivated by Smak's inferred initial disk mass $\sim10^{24}$ g.
 Figure 2 shows the light curves and associated decay
time scales for these models.
As expected, the deviations from exponentiality are seen here
as well. It is interesting that the dynamic range over
which $\tau(m_V)$ changes is 
      about a factor of two, which
is  slightly less than seen in WZ Sge.
This may be a reflection of the crude (Planckian)
flux distributions which are used to  model the local
emissivity in the disk.
 The calculations of $\tau$ based on ${\dot M}_{\rm inner}$
show a larger dynamic variation.

\section {Discussion }

Our result supports the standard model
  for dwarf nova outbursts in which
the outbursting disk decays primarily
via simple accretion onto the WD when 
in the ``viscous plateau'' (Cannizzo 1993b) 
 stage, and $\alpha_{\rm hot}\simeq 0.1-0.2$.
The faster decay which began 18 Aug 2001
in WZ Sge would then be caused by the onset
of the cooling front,
its rapidness in comparison to the previous
evolution stemming from the fact
that the thermal time scale is shorter
than the viscous time scale.
It is fortunate that nature has given us
the WZ Sge stars which remain in a viscous 
plateau long enough for one to be able
to constrain the functional form of the decay
  as being demonstrably different from exponential.
For longer period dwarf novae such as SS Cyg
($P_{\rm orb}=6.6$ hr)
in which 
  one frequently sees flat-topped outbursts
 where the disk is also 
  presumably 
    entirely in the hot state,
the short duration ($\sim7-10$ d) 
 of the viscous state due to a smaller
   initial 
surface density excess $\Sigma(r_{\rm outer})/\Sigma_{\rm min}(r_{\rm outer})$
in the outer disk,  combined
  with the 
  slower viscous time at the outer
disk (due to larger $r_{\rm outer}$),
conspire to produce a much smaller
dynamic variation in $m_V$.
In fact, the variations that are seen may be
due almost entirely 
      to sloshing action of the gas in
the disk as it responds
to the matter redistribution accompanying the outburst.

It is interesting to ponder the implications
for  other types of systems, 
for example the soft X-ray transients (SXTs)
     in which the
accreting object is a neutron star or black hole.
The similarity between the
superoutbursts in the WZ Sge stars
and the  X-ray nova outbursts seen  in the 
  some of the brightest and best studied
   SXTs
has been pointed out by
 Kuulkers et al. 
(1996, see also Kuulkers 1998).
King \& Ritter (1998, see also King 1998, 
     Shahbaz, Charles, \& King
1998)
     propose
       that the $\sim30-50$ d $e-$folding
decay times seen in these SXTs are viscous,
insomuch
   as one expects strong irradiation in the outbursting
state to prevent the cooling front from forming.
Cannizzo (2000) argues that
  even though irradiation keeps the entire
  disk ionized,
    the viscous time scale
 would be too slow
(the time dependent  computations
 of Dubus et al. [2001, e.g. Fig. 16a]
 show an $e-$folding
    decay  time of $\sim 100-200$ d during
 the viscous plateaus of their outbursts),
%
and Cannizzo 
      hypothesizes that some other agent
 such as strong evaporation must be
at work in outburst to reduce the disk mass
from the inner disk
on a time scale shorter
than that due to accretion acting alone.

A scaling of the viscous time at the outer
edge applied to A0620-00,
which had a bright outburst
and exponential decay in 1975,
     gives

$\tau_v  \simeq 400$ d
$(M_1/10\msun)^{5/14}$  $r_{11}^{13/14}$  $(\alpha_{\rm hot}/0.1)^{-8/7}$
$(\Sigma/ 10^2$ g cm$^{-2})^{-3/7}$.

\noindent Such a long time is problematic
in accounting for the $\sim30-50$ d
$e-$folding decay times seen in systems
with a variety of orbital periods
and hence outer disk radii
(see Table 1 of Mineshige, Yamasaki, Ishizaka 1993),
and more importantly, the rise times
     of $\sim1-3$ d
for the secondary
maxima in the SXTs
which, in the model of King \& Ritter (1998),
are due to the arrival of matter 
from the outer disk to the inner disk.
In reality, matter added to the disk at
the outer edge
and diffusing inward
would spread out and not produce
such a short, well-defined rise in
the X-ray light curve.
Furthermore, it would be
difficult to maintain an exponential
decay over a dynamic range of
$\sim10^2 - 10^3$ in $L_X$ as observed,
due to the decrease in $\Sigma(r_{\rm outer})$
and the fact that $\tau_v\propto \Sigma(r_{\rm outer})^{-3/7}$.
  (The dynamic ranges for viscous decays shown in Dubus et al.
2001
      are only about a factor of 10.)

Could irradiation of the outer disk
    in the SXTs
     affect
the midplane temperature enough to
change significantly the viscous time scale?
  For irradiation to be
strong enough to have a significant effect on the midplane
temperature (which enters into the disk semithickness and
  hence viscous time
scale), the local irradiation temperature needs to exceed
not just the local non-irradiated disk photospheric temperature, but
also the midplane temperature which is $\sim\tau^{1/4}$
     times the effective
temperature, where $\tau$ is the vertically integrated optical depth.
Starting again with eqn. [A6] from 
Cannizzo \& Reiff (1992), one can show

$\tau = \kappa \Sigma \simeq 3000  (M_1/10 \msun)^{-1/14}
        r_{11}^{3/14}
               (\alpha_{\rm hot}/0.1)^{-4/7}
         (\Sigma/100 \ {\rm g} \  {\rm cm}^{-2})^{2/7}$,

\noindent an optical depth of $\sim10^3 - 10^4$ 
      through the vertical structure,
so that one would need a local irradiation temperature about an order
of magnitude greater than the effective temperature. 
     Realistic calculations
of the irradiation effect in SXTs show  that
the change in midplane temperature is small (Dubus et al. 1999). 
 In other words, irradiation indeed keeps the entire disk
ionized
  during the ``viscous'' decay
(because the irradiation temperature
exceeds $\sim10^4$ K at the outer disk edge),
     but because of the large 
   optical depth one does not see a
significant change in midplane temperature.

\section{ Conclusion }

We have examined the 2001 superoutburst
light curve of WZ Sge.
We find that the deviation from exponentiality
during the time the system was in superoutburst
is roughly characterized
by a linear increase in the local
decay time from $\sim4$ d mag$^{-1}$
       to $\sim12$ d mag$^{-1}$
  up to about two-thirds of the way
through superoutburst.
 This is consistent with standard limit cycle
accretion disk theory, adopting systemic parameters
for WZ Sge taken from Smak (1993),
namely an initial disk mass $10^{24}$ g,
a central mass $0.45\msun$, an outer disk radius
$10^{10}$ cm, and an alpha value $\alpha_{\rm hot} = 0.1$.
Among the dwarf novae, only the WZ Sge stars
show a large enough dynamic range in $m_V$ in
their stages of viscous decay during outburst
to allow this type of study.

An application of the viscous time $\tau_v(\alpha_{\rm hot},
r_{\rm outer},\Sigma)$
scaling law to outbursts
in the soft X-ray transients appears to lead
to decay times which are slower than observed
in several of the bright, well-studied systems
(see 
    Fig. 9 of Kuulkers 1998).
Also, the rise times of $\sim1-3$ d for the secondary
maxima seen in these systems
are especially
      difficult to reconcile with the class of theories
in which material added at the outer disk edge
must diffuse to the inner edge (Chen et al. 1993,
Augusteijn et al. 1993, King \& Ritter 1998)
 since the diffusion time and the smearing time
would both be $\sim\tau_v(r_{\rm outer})$.

The data were obtained from the VSNET website
{ \tt www.kusastro.kyoto-u.ac.jp/vsnet/etc/searchobs.cgi?text=SGEWZ}.
We thank Erik Kuulkers for pointing us to this link.
  We also thank Tom Marsh for organizing
a stimulating {\it ad hoc} session on WZ Sge
at the August 2001 CV meeting in 
    Goettingen, Germany
(``The Physics of Cataclysmic Variables and Related Objects'').

\vfil\eject
\centerline{ FIGURE CAPTIONS }

Figure 1. The amateur $m_V$ data from the VSNET 
website for the July-August 2001 superoutburst
of WZ Sge ({\it top panel}),
the data binned into 0.2 d bins and smoothed
  in 2 d moving averages ({\it middle panel}),
and the decay time scale associated with 
the smoothed light curve ({\it bottom panel}).
  The two straight lines in the second panel 
indicate a decay time of 10 d mag$^{-1}$.

Figure 2. The model light curve for 
the outburst described in the text, using
parameters relevant for WZ Sge.
Shown are the $M_V$ light curve ({\it top panel}),
the disk mass ({\it middle panel}),
and the locally defined decay time scale
along the light curve ({\it bottom panel}).
  The two curves given in each panel 
are for an initial disk mass of $10^{24}$ g
and $3\times 10^{24}$ g,
with the latter curve extending further to the
right.
  The two straight lines in the top panel
indicate a decay time of 10 d mag$^{-1}$.
The additional two sets of curves which appear toward
the bottom of the third panel show the 
decay time scale computed not from the $M_V$
light curve but rather from the rate of mass loss
from the inner edge onto the WD.
  It appears that the effect of computing
a $V$ band light curve introduces a flattening
into the deviation from exponentiality: the 
change in $\tau$ in the model light curve
is only about a factor of two, versus about a 
factor of three for the observed WZ Sge superoutburst
light curve.

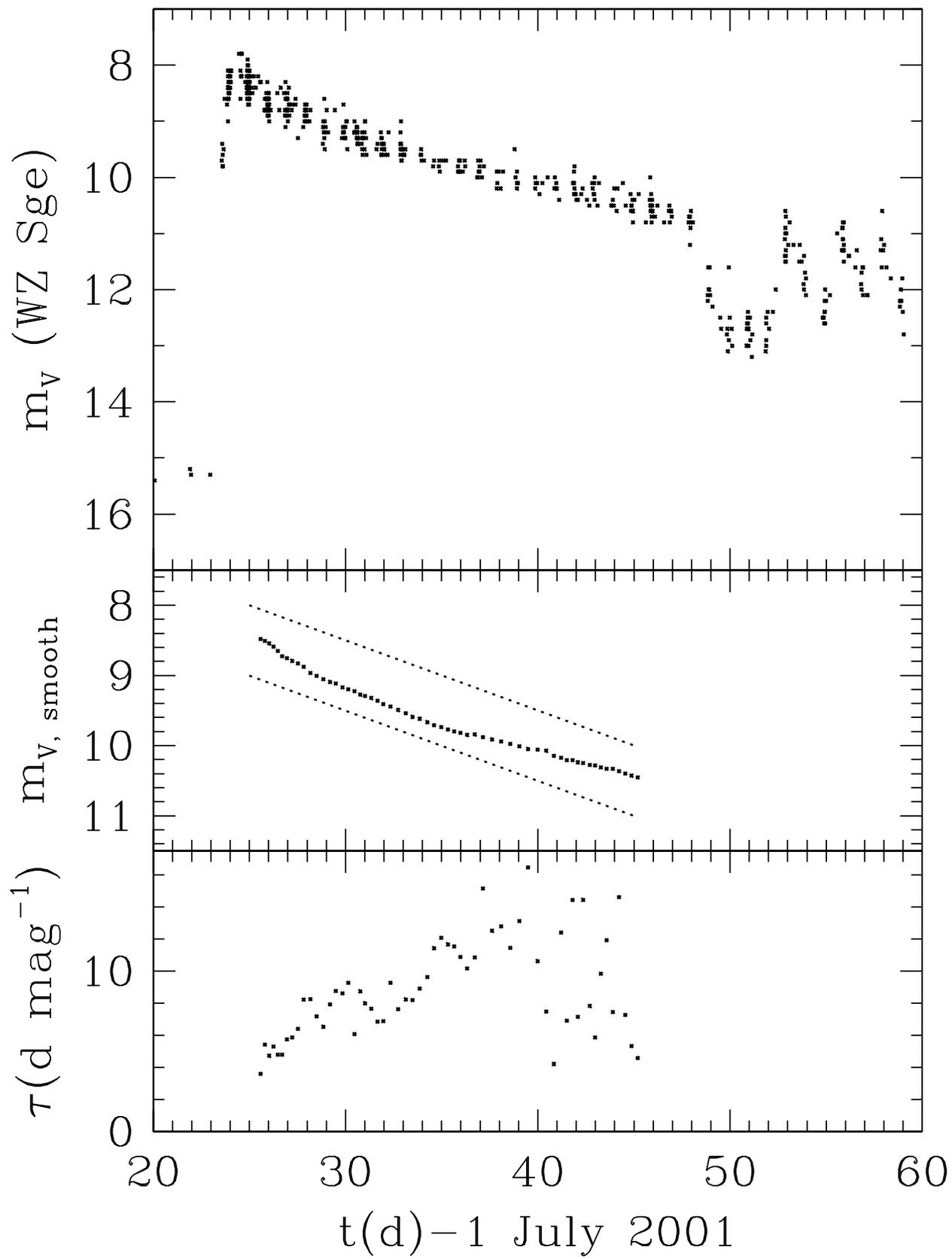
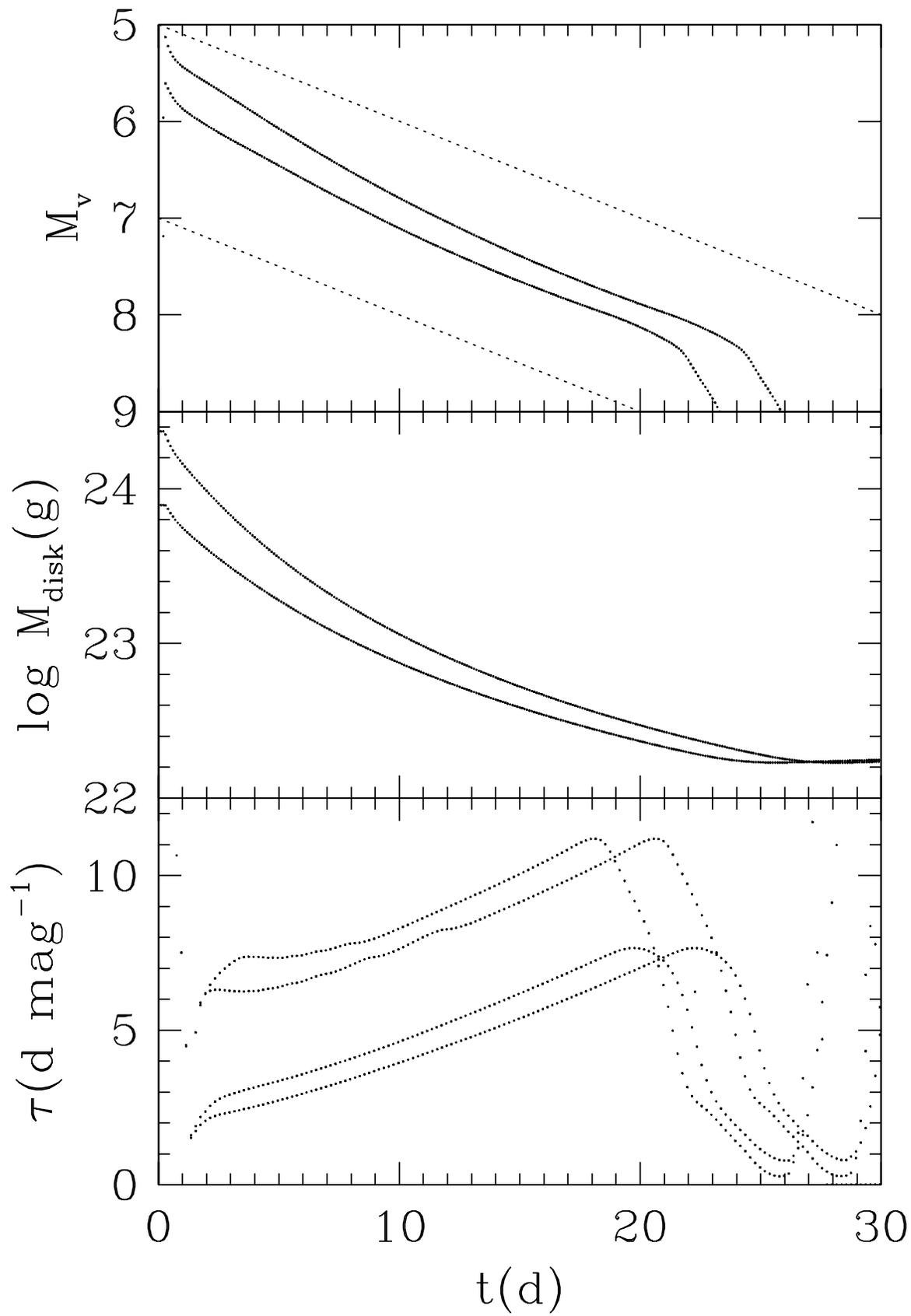
\end{document}